\documentclass{knawproc}

\def\spose#1{\hbox to 0pt{#1\hss}}

\def\lya{\ifmmode {\rm\,Ly\alpha}\else ${\rm\,Ly\alpha}$\fi}
\def\Mdot {\ifmmode {\rm {\dot M}} \else ${\rm {\dot M}}$\fi}

\def\kms{\ifmmode {\rm\,km\,s^{-1}}\else
    ${\rm\,km\,s^{-1}}$\fi}
\def\kmsMpc{\ifmmode {\rm\,km\,s^{-1}\,Mpc^{-1}}\else
    ${\rm\,km\,s^{-1}\,Mpc^{-1}}$\fi}

\def\msun{\ifmmode {\rm\,M_\odot}\else ${\rm\,M_\odot}$\fi}
\def\Msun{\ifmmode {\rm\,M_\odot}\else ${\rm\,M_\odot}$\fi}
\def\lsun{\ifmmode {\rm\,L_\odot}\else ${\rm\,L_\odot}$\fi}
\def\Lsun{\ifmmode {\rm\,L_\odot}\else ${\rm\,L_\odot}$\fi}
\def\rsun{\ifmmode {\rm\,R_\odot}\else ${\rm\,R_\odot}$\fi}
\def\Rsun{\ifmmode {\rm\,R_\odot}\else ${\rm\,R_\odot}$\fi}

\def\cm{{\rm\,cm}}
\def\cm3{\ifmmode {\rm\,cm^{-3}}\else ${\rm\,cm^{-3}}$\fi}

\def\ergps{\ifmmode {\rm\,erg\,s^{-1}}\else ${\rm\,erg\,s^{-1}}$\fi}
\def\ergpscm2{\ifmmode {\rm\,erg\,s^{-1}\,cm^{-2}}\else
    ${\rm\,erg\,s^{-1}\,cm^{-2}}$\fi}

\def\eg{{e.g.}}
\def\deg{\ifmmode {^{\circ}}\else {$^\circ$}\fi}
\def\degr{\ifmmode {^{\circ}}\else {$^\circ$}\fi}
\def\degs{\ifmmode {^{\circ}}\else {$^\circ$}\fi}

\def\etal{{et al.~}}

\def\h3Mpc{h^{-3}{\rm Mpc}^3}
\def\Ho{\ifmmode {\rm\,H_0}\else ${\rm\,H_0}$\fi}
\def\hnot{\ifmmode {\rm\,H_0}\else ${\rm\,H_0}$\fi}
\def\h0{\ifmmode {\rm\,H_0}\else ${\rm\,H_0}$\fi}
\def\hnotunit{\ifmmode {\rm\,km\,s^{-1}\,Mpc^{-1}}\else
    ${\rm\,km\,s^{-1}\,Mpc^{-1}}$\fi}
\def\qnot{\ifmmode {\rm\,q_0}\else ${\rm q_0}$\fi}
\def\q0{\ifmmode {\rm\,q_0}\else ${\rm q_0}$\fi}
\def\ie{{i.e.}}
\def\mic{\ifmmode {\rm\,\mu m}\else ${\rm \mu m}$\fi}

\def\arcsec{\ifmmode {^{\prime\prime}}\else $^{\prime\prime}$\fi}
\def\asec{\ifmmode {^{\prime\prime}}\else $^{\prime\prime}$\fi}
\def\arcmin{\ifmmode {^{\prime}}\else $^{\prime}$\fi}
\def\amin{\ifmmode {^{\prime}}\else $^{\prime}$\fi}

\def\secper{\ifmmode \rlap.{^{s}}\else $\rlap{.}{^{s}} $\fi}
\def\minper{\ifmmode \rlap.{^{m}}\else $\rlap{.}{^m} $\fi}
\def\magper{\ifmmode \rlap.{^{m}}\else $\rlap{.}{^m} $\fi}
\def\farcs{\ifmmode \rlap.{^{\prime\prime}}\else
    $\rlap.{^{\prime\prime}}$\fi}
\def\arcsper{\ifmmode \rlap.{^{\prime\prime}}\else
    $\rlap.{^{\prime\prime}}$\fi}
\def\arcmper{\ifmmode \rlap.{^{\prime}}\else
    $\rlap.{^{\prime}}$\fi}
\def\spose#1{\hbox to 0pt{#1\hss}}
\def\simlt{\mathrel{\spose{\lower 3pt\hbox{$\mathchar"218$}}
     \raise 2.0pt\hbox{$\mathchar"13C$}}}
\def\simgt{\mathrel{\spose{\lower 3pt\hbox{$\mathchar"218$}}
     \raise 2.0pt\hbox{$\mathchar"13E$}}}

\def\araa{{ARA\&A}}
\def\aa{{A\&A}}

\def\apj{{ApJ}}
\def\apjlett{{ApJ}}

\def\mn{{MNRAS}}
\def\mnras{{MNRAS}}
\def\nature{{Nature}}

\def\apjref#1;#2;#3;#4 {\par\pp#1, {#2}, #3, #4 \par}

\def\plotfiddle#1#2#3#4#5#6#7{\centering \leavevmode
\vbox to#2{\rule{0pt}{#2}}
\includegraphics{#1}}

\begin{document}

\begin{opening}

\title{The Early History of Powerful Radio Galaxies}

\author{Arjun Dey}
\addresses{Kitt Peak National Observatory, 950 N. Cherry Ave., P. O. Box 26732, Tucson, AZ 85726-6732, USA\\
Dept.~of Physics \& Astronomy, The Johns Hopkins University, Baltimore, MD 21218, USA
}
\runningtitle{High-z RGs}
\runningauthor{Arjun Dey}

\end{opening}


\begin{abstract}

I briefly review the current status of observations of AGN-powered
UV/optical light, starlight, dust and outflow phenomena in
high-redshift powerful radio galaxies. The existing data are consistent
with the hypothesis that powerful radio galaxies undergo a major episode
of star formation at high redshift ($z\simgt 4$) during which they
form most of their stars, and subsequently evolve `passively', with the
UV continuum emission in the $z\sim 1$ galaxies being dominated by
AGN-related processes rather than starlight from the underlying, aging
population.

\end{abstract}


\section{Introduction}

The host galaxies of distant radio sources are of fundamental
importance to studies of galaxy formation and evolution primarily
because these objects are the best candidates for the progenitors of
present-day massive galaxies and represent strongly biased peaks in the 
matter distribution. This hypothesis is supported at low
redshift by the association of powerful radio sources with gE and cD
galaxies (Matthews \etal 1964), at intermediate and high redshifts by
the tendency for these sources to reside in moderately rich cluster
environments (\eg, Hill \& Lilly 1991, Dickinson 1997), and by a few
direct kinematic measurements of the masses of high-redshift powerful
radio galaxies (\eg, 3C265: Dey \& Spinrad 1996). In this contribution,
I will briefly review the state of the observational data on
high-redshift powerful radio galaxies (hereafter HzRGs) and their
relevance to our understanding of the evolution of these systems.

\begin{figure}[t]
\plotfiddle{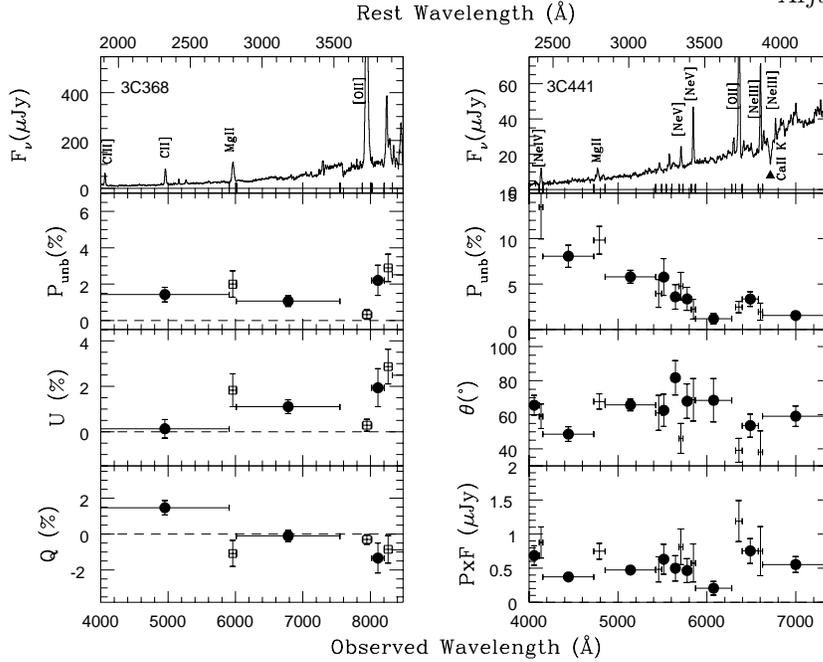}{2.8in}{-90}{47}{47}{-190}{250}
\caption{Spectropolarimetric observations of two radio galaxies: 3C368 at $z=1.132$ (left panel) and 3C441 at $z=0.707$ (right panel).
Since 3C368 is unpolarized, with $P<3\%$, 
only the Stokes parameters ($Q$ and $U$) and the unbiased
percentage polarization ($P_{unb}$) are shown.  In contrast, 3C441 is
strongly polarized and shows a monotonically decreasing $P$ and
wavelength-independent polarized flux. Note the CaII~K stellar absorption in
3C441, and the polarization of the MgII$\lambda$2800 emission line.}
\label{3c441pol}
\end{figure}

\section{AGN--Powered UV and Optical Light}

A significant fraction of the observed optical (rest-frame UV) emission
from $z\sim 1-2$ radio galaxies is now known to be non-stellar, and
dominated by scattered and reprocessed light from the active nucleus. A
connection between the AGN and the UV continuum and line emission from
HzRGs is clearly evidenced by the `alignment effect', the tendency for
the UV emission to be spatially extended along the major axis of the
radio emission (McCarthy \etal 1987; Chambers \etal 1987).  Several
imaging- and spectro-polarimetric studies (Tadhunter \etal 1988; di
Serego Alighieri \etal 1989, Jannuzi \& Elston 1991, Jannuzi \etal
1995, Dey \etal 1996, Cimatti \etal 1996, 1997) have now convincingly
demonstrated that the extended, aligned UV continuum emission in most
$z\sim 1-2$ HzRGs is strongly polarized, with the electric vector
oriented perpendicular to the major axis of the UV emission. These
observations provide strong evidence that much of the UV light is
anisotropically radiated emission from the AGN scattered into our line
of sight by dust and electrons in the ambient medium.

Our observations of 3C441 (shown in the right panel of
figure~\ref{3c441pol}) are fairly typical of the $z\sim 1-2$ radio
galaxies: the fractional polarization ($P$) is large ($\simgt$5\% at
$\lambda<3200$\AA) and blue (\ie, decreasing with increasing
wavelength), but the  polarized flux ($P\times F$) is roughly
wavelength--independent. Not all powerful radio galaxies at $z\sim 1-2$
are polarized.  The aligned radio galaxy 3C368 at $z=1.132$ remains a
notable exception (left panel of figure~\ref{3c441pol} \& table 1);
however, even in this case, the extended UV continuum emission (at rest
wavelengths $\lambda$2000--3500\AA) is largely accounted for by
optically thin Balmer continuum emission from the line-emitting gas
(Stockton \etal 1996) powered by the AGN.

In a few cases where the data are of high enough signal-to-noise ratio
(\eg, 3C265 --- Dey \& Spinrad 1996, Tran \etal 1998; 3C324 --- Cimatti
\etal 1996; 3C441 --- figure~\ref{3c441pol}; 3C277.1 - Tran \etal 1998),
there is some evidence that the hidden nuclear source has quasar-like
broad emission lines.  It is important to note, however, that the total
number of $z\simgt 1$ radio galaxies for which high qualtity
data exist is small (the observations require
long exposures of $\simgt 1$ night per object even on the largest
telescopes) and hence concluding that {\it all} powerful radio galaxies
contain quasars hidden from direct view is premature.

At higher redshifts ($z>3$), the data are even more sparse: polarimetry
exist for only two $z>3$ powerful radio galaxies as of this writing,
and both of these are unpolarized (see table). Instead, the spectra of
these galaxies show strong resonance absorption features typically
observed in nearby starburst galaxies (\S~3). At least in the case of
4C41.17 ($z=3.8$; Chambers \etal 1990), the lack of polarization, the
absorption lines, and the inadequacy of alternative emission processes
(e.g., Balmer continuum emission) to account for the observed extended
UV continuum emission together imply that starlight dominates the UV
light from this system (Dey \etal 1998).

\begin{table}[t]
\begin{tabular}{lccrc}
\hline
\hline
{\bf Name} & {\bf Redshift} & {\bf Rest} & {\bf \%\ \ } & {\bf \ \ \ References} \\
&  & {\bf Wavelength} & {\bf Polzn}  &      \\
\hline
\hline
\underline{$z\sim 0.7-2$}  & & & & \\
\ \ 3C441     & 0.706 & 2430-2770~\AA\ &  8\% & 6 \\
\ \ 3C343.1   & 0.750 & 2857-4343~\AA\ & $<$2.2\% & 12 \\
\ \ 3C277.2   & 0.766 & 2325-2893~\AA\ & 29\% & 7, 12 \\
\ \ 3C265     & 0.811 & 2200-2500~\AA\ & 14\% & 9, 12 \\
\ \ 3C226     & 0.818 & 2000-2700~\AA\ & 13\% & 8 \\
\ \ 3C356a    & 1.079 & 1930-2310~\AA\ & 14\% & 2 \\
\ \ 3C368     & 1.132 & 1880-2350~\AA\ &$<$1\%& 6 \\
\ \ 3C324     & 1.206 & 1915-2120~\AA\ & 12\% & 1 \\
\ \ 3C13      & 1.351 & 1701-2300~\AA\ &  7\% & 2 \\
\ \ 3C256     & 1.819 & 1420-1540~\AA\ & 10\% & 4, 10 \\[1ex]
\underline{$z\sim 2-3$} &  & & & \\
\ \ 4C00.54   & 2.37  & 1250-1530~\AA\ & 12\% & 3 \\
\ \ 4C23.56a  & 2.419 & 1250-1520~\AA\ & 15\% & 3 \\
\ \ TX0830+191 & 2.572 & 1550-2140~\AA\ & 13\% & 11 \\[1ex]
\underline{$z\simgt 3$} & & & & \\
\ \ 6C1909+722& 3.534 & 1270-1520~\AA\ & $<$5\%   & 6 \\
\ \ 4C41.17   & 3.800 & 1220-1540~\AA\ & $<$2.4\% & 5 \\[1ex]
\hline
\hline
\end{tabular}

{\small
Key to references: 
[1] Cimatti \etal (1996);
[2] Cimatti \etal (1997);
[3] Cimatti \etal (1998);
[4] Dey \etal (1996);
[5] Dey \etal (1997);
[6] Dey \etal (1998);
[7] di Serego Alighieri \etal (1989);
[8] di Serego Alighieri (1997);
[9] Jannuzi \& Elston (1991);
[10] Jannuzi \etal (1995);
[11] Knopp \& Chambers (1997);
[12] Tran \etal (1998)
}
\end{table}

An important question is whether or not there is any evidence for
evolution in the fractional AGN contribution to the UV continuum
emission in high redshift radio galaxies. Although we remain in the
preliminary stages of investigating this problem, the existing data
(see table), albeit limited, hint at a decrease in $P$ with
increasing redshift.  One has to be cautious in interpreting the
current data: at different redshifts, the current observations sample
different rest wavelength ranges, and the observed differences with
redshift could result from a wavelength dependent emergent fractional
polarization (either due to the intrinsic nature of the scattering
process or to dilution by an unpolarized component) rather than an
evolutionary phenomenon. The existing observations only permit us to
compare the same rest wavelength range in a few objects 
in the redshift range where evolution may be expected. Clearly,
optical polarization observations of objects (with similar radio
luminosities) in the intermediate redshift range between 2 and 3 and
near-IR polarimetry of the $z>3$ galaxies will provide the critical
tests of the reality of any evolution in $P$ (\eg, Cimatti \etal 1998).

However, if the decrease in $P$ (at a fixed rest wavelength) with
redshift proves to be real, it is most likely to be due to a variation
in the relative dominance of the scattered, polarized component over
the diluting, unpolarized components at rest-frame UV wavelengths. 
An evolutionary 
picture in which the unpolarized component decreases in intensity
relative to the polarized component is consistent with the present
observations, and may be expected if the diluting component is
dominated by starlight from an aging population. For instance, the time
elapsed between $z=3.8$ and $z=2.5$ is nearly 1.2~Gyr (\hnot=50,
\qnot=0.1, $\Lambda$=0), and would provide ample time for a starburst
to age sufficiently so as to contribute negligible flux to the UV
spectrum.  This hypothesis is also supported by the evidence for stars
in HzRGs, which is discussed in the next
section, and provides a natural explanation for the uniformity observed
in the $K-z$ Hubble diagram (\eg, Lilly \& Longair 1984).

\section{Starlight}

\begin{figure}[t]
\plotfiddle{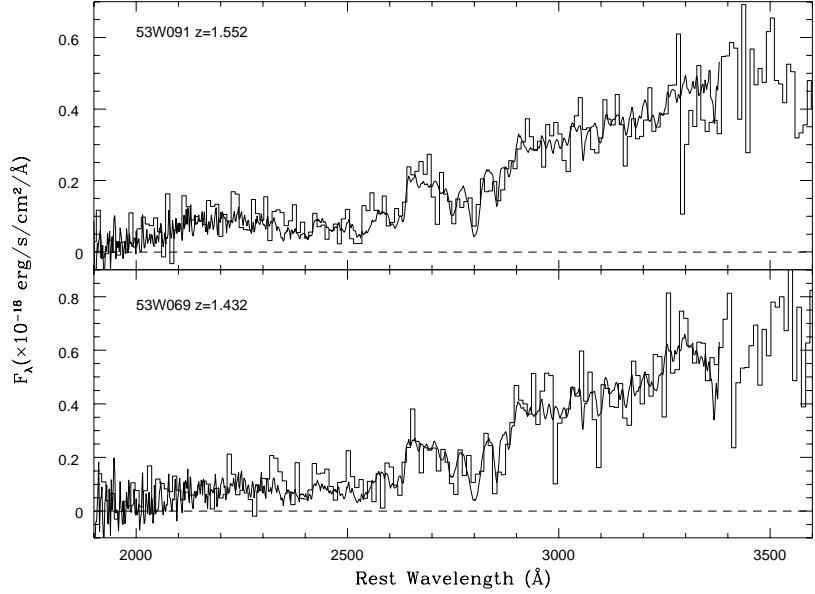}{2.7in}{-90}{42}{42}{-165}{240}
\caption{Spectra of 
LBDS~53W091 and LBDS~53W069 overplotted with the {\it IUE} spectra of 
late F-type main sequence stars.} 
\label{lbdsspec}
\end{figure}

Until recently, the arguments for the presence of stars in distant
radio galaxies were based primarily on near-IR observations, which
showed symmetric rest-frame optical morphologies (similar to normal
elliptical galaxies) in contrast to the peculiar, elongated,
multi-component rest-frame UV structures observed at optical
wavelengths (McCarthy \etal 1987, Chambers \etal 1987, Eisenhardt \&
Chokshi 1990, Rigler \etal 1992, Dunlop \& Peacock 1993; see McCarthy
1993 for a review). In addition, the landmark study of Lilly \& Longair
(1984) demonstrated that the photometric properties of HzRGs 
form a homogeneous redshift sequence (the
`$K$-$z$ Hubble diagram'), which is well-modeled by the `passive'
evolution of a stellar population formed at high redshift (\eg, McCarthy
1993). The small scatter in the $K$-$z$ diagram exhibited by the host
galaxies of at least the powerful radio sources has been generally
taken to suggest that these objects are the progenitors of the
present-day giant ellipticals and cD galaxies.  However, given the
faintness of the distant radio galaxies and the limitations of current
near-IR spectrographs, {\em direct} spectroscopic evidence for
starlight in HzRGs has proved elusive.

\begin{figure}[t]
\plotfiddle{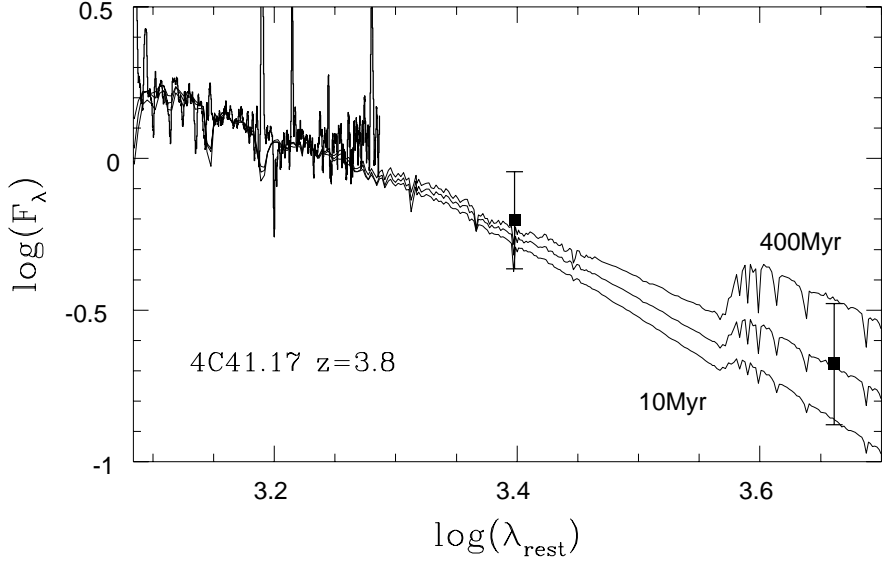}{2.7in}{0}{70}{70}{-200}{-165}
\caption{Population synthesis model fits to the spectrum of 4C41.17.
The curves show Bruzual \& Charlot (1996) constant star-forming models
(assuming a Salpeter IMF $1<M<100\Msun$, $Z_\odot$, and reddened by
$E_{B-V}=0.1$) at ages of $10^7$, $7.2\times10^7$, and $4\times10^8$ yrs, 
superimposed on the rest-frame UV spectrum and optical
broad-band colours. The models are normalized to the spectra at $\lambda_{rest}=1500$\AA.}
\label{4cage} 
\end{figure}

High signal-to-noise ratio, moderate resolution spectroscopic studies
have become possible using the new, large aperture telescopes. Our
group's spectropolarimetric and spectroscopic observations at the
W.~M.~Keck Observatory have resulted in the detection of the CaII~K
absorption line in several HzRGs (\eg, Dey \& Spinrad 1996; also right
panel of figure~\ref{3c441pol}); the strength of this line suggests
that it arises in stellar photospheres rather than in the interstellar
media of the host galaxies.  Stockton \etal (1996) have presented deep
Keck spectroscopy of 3C65 ($z=1.175$) and attempted to subtract the
strong emission lines and then fit a population synthesis model to the
underlying continuum spectrum.

The strongest evidence to date that at least some HzRGs
contain stars is provided by the spectra of three objects: LBDS~53W091
at $z=1.55$ (Dunlop \etal 1996, Spinrad \etal 1997), LBDS~53W069 at
$z=1.43$ (Dey \etal 1998), and 4C41.17 at $z=3.8$ (Dey \etal 1997, Chambers
\etal 1990).  The first two objects are radio galaxies which have lower
radio luminosities than similar redshift 3CR galaxies, but are,
nevertheless, powerful radio sources (\ie, more luminous than the break
in the radio luminosity function).  The last, 4C41.17, is a powerful
radio galaxy which exhibits an unpolarized UV continuum and absorption
lines similar to those observed in nearby starbursts.

The two LBDS radio galaxies have very weak (insignificant) AGN
contributions at rest-frame UV wavelengths, and their spectra at these
wavelengths show unmistakable signatures of an old stellar population.
In fact, the spectra of both LBDS galaxies are well represented by the
{\it IUE} spectra of individual F-type main sequence stars
(figure~\ref{lbdsspec}).  Since the rest-frame UV light in a simple
instantaneous burst stellar population is dominated by the main
sequence turn-off stars for ages between 1 and 7 Gyr (\eg, Bruzual
\& Charlot 1996, Spinrad \etal 1997), the implied colour of the
turn-off can be used to place a stringent lower limit to the age of the
galaxy. The implied rest-frame $B-V$ colors of the two galaxies can be
used in conjunction with the Revised Yale Isochrones to determine a
lower limit to the age of the stellar population that dominates the UV
light as a function of metallicity.  More sophisticated population
synthesis modelling (Dunlop \etal 1996, Spinrad \etal 1997, Dunlop
1998) reinforce this simple analysis, and provide age estimates of
between 3 -- 5~Gyr (for relatively normal IMFs) for the stellar
populations in the two LBDS galaxies.

A similarly detailed analysis is difficult to perform for the more
luminous 3CR radio galaxies, since the spectrum of the UV starlight
is diluted by strong emission lines and nonstellar continua (\eg,
scattered light and Balmer continuum emission). However, the symmetric,
`elliptical' near-IR morphologies of most $z\sim 1-2$ 3CR galaxies
suggests that their stellar components are also dynamically old (Rigler
\etal 1992, Best \etal 1997). The similarity in $K$-band morphologies
of the LBDS and 3CR galaxies may provide weak support for generalizing
the results derived for LBDS~53W069 and LBDS~53W091 to the more luminous
3CR systems, suggesting that most luminous radio galaxies at $z\sim 1.5$
contain evolved populations as old as 3 -- 5~Gyr. These large ages imply
a high formation redshift ($z_f$) for the dominant UV population of the
$z\sim 1-2$ powerful radio galaxies, and may also pose stringent lower
limits on the age of the Universe. For instance, the age of 4~Gyr 
for LBDS~53W069 derived using population synthesis techniques implies 
a $z_f > 5$ ($\Omega=0.2$).

These high formation redshifts for the $z\sim 1-2$ powerful radio
galaxies imply that their counterparts at $z\sim 4$ should be observed
to contain stellar populations a few hundred million years old. It is
therefore very intriguing that 4C41.17, the only $z\sim 4$ powerful
radio galaxy for which sufficiently detailed data permit a fairly
reliable age estimate, has a population not older than a few hundred Myr (Dey
\etal 1997; see also figure~\ref{4cage}). The star-formation rate
derived for 4C41.17 from its rest-frame 1500\AA\ luminosity is $\sim
140 - 1100~\Msun~{\rm yr}^{-1}$, and a factor of 3 larger if the galaxy is
reddened by $E_{B-V}\sim 0.1$. Sustained for a period of order 0.5~Gyr,
this rate will produce a gE's equivalent of stars. 

\section{Dust}

There are now several lines of evidence that point to the existence of
significant quantities of dust in HzRGs. At
least four $z>2$ objects (8C1435+635 at $z=4.25$ -- Ivison \etal 1997;
4C41.17 at $z=3.8$ -- Dunlop \etal 1994; MG1019+0535 at $z=2.8$ -- Cimatti
\etal 1998b; 53W002 at $z=2.4$ -- Hughes \etal 1997) have robust
detections of sub-mm continuum flux in excess of the expected
nonthermal component (extrapolated from radio wavelengths).  In
addition to this direct evidence, the presence of dust in HzRGs can
also be inferred by the reddening of the line emission and starlight
and from the measured UV / optical polarization. For instance, the
stellar components in both 3C324 and 3C368 are clearly observed at
near-IR wavelengths (\ie, rest-frame optical $\lambda$s), but are
invisible in the deep HST WFPC2 imaging data (rest-frame UV
$\lambda$s); the upper limits suggest dust reddening in excess of
$E_{B-V}\sim 0.3$ (\eg, Dickinson, Dey \& Spinrad 1995). Finally, as
discussed in \S~2, the observed UV / optical polarization measurements
are generally interpreted in terms of scattering of anisotropically
radiated AGN light, and dust particles can provide an efficient
scattering population. However, since the scattering and polarizing
properties of dust grains depend on their chemistry, size distribution,
optical depth and spatial distribution, at present it is only possible
to derive very crude constraints on the global properties of the
scattering population from the polarization observations.  These
observations suggest that there is a large quantity of dust which is
distributed over a fairly large volume which scatters and reddens the
observed optical continuum emission. Estimates of the dust mass range
from $M_{\rm dust}\sim 0.2 - 2 \times 10^8$ \Msun\ derived from the
sub-mm continuum emission to $\sim 10^5 - 10^6$~\Msun\ derived from
reddening / scattering data. The sub-mm data also provide
crude estimates of the effective temperature of the dust which range
between 30 -- 200 K (Hughes \etal 1998, Ivison \etal 1998, Cimatti \etal 1998).

The existence of large amounts of dust in the highest redshift radio
galaxies known poses a puzzle: since dust formation is predicated on
the existence of metals (and therefore on star-formation), how did so
much dust form in such a short time from the Big Bang? Very little is
known about the formation and evolution time-scales of dust, especially
under the extreme conditions that may prevail in ISM of HzRGs. However, it
is intriguing to note that simply scaling the dust injection rate due
to supernovae inferred for our Galaxy (\eg, Whittet 1992) by the
star-formation rate observed in the the high-redshift systems can account for
in the large amounts of dust observed ($M_{\rm dust} \sim 7\times 10^8
[\Mdot/10^3\Msun~{\rm yr}^{-1}][t/1{\rm Gyr}]$~\Msun; where \Mdot\ is the
star-formation rate and $t$ is the duration of the starburst).  Another 
important question concerns the ultimate fate of the dust
grains.  Low redshift cD and gE galaxies, which are thought to be the
evolutionary products of the HzRGs, do not
contain large quantities of dust. One way to efficiently destroy dust
may be through sputtering processes in the shocks associated with 
the radio source (DeYoung 1998).

\begin{figure}[h]
\plotfiddle{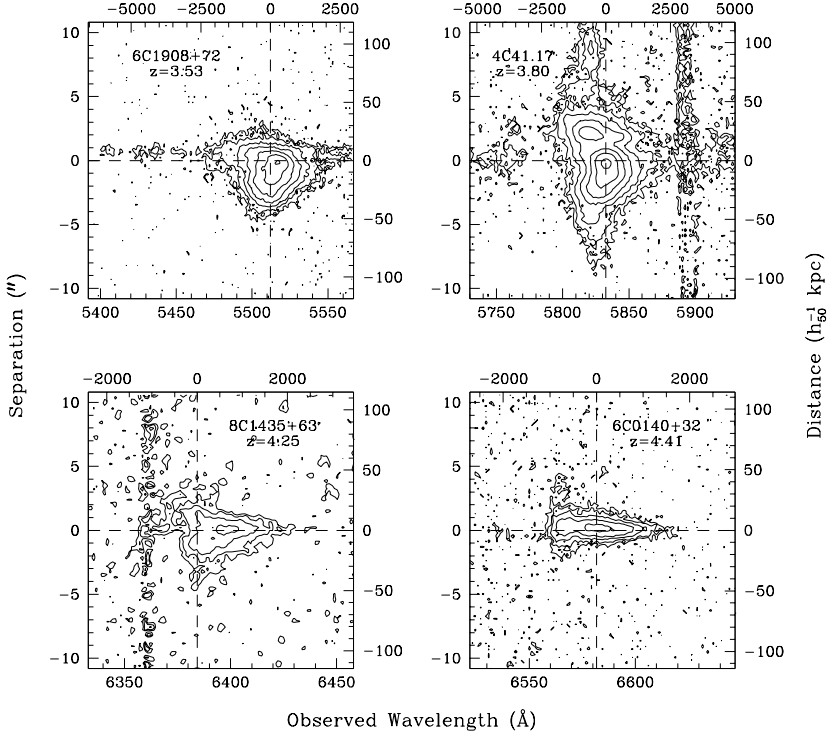}{3.3in}{0}{58}{58}{-180}{-110}
\caption{Ly$\alpha$ emission profiles of four HzRGs.
The triangular-shaped profiles result from spatially extended absorption
which is blue-shifted relative to the systemic velocity. The zero velocity
in all cases is determined either from the HeII$\lambda$1640 line or the
CIII{]}$\lambda$1909 line. Note that the \lya\ emission from 4C41.17 
extends over $>200$~kpc.} \label{lyapanels}
\end{figure}

\section{Outflows}

\begin{figure}[t]
\plotfiddle{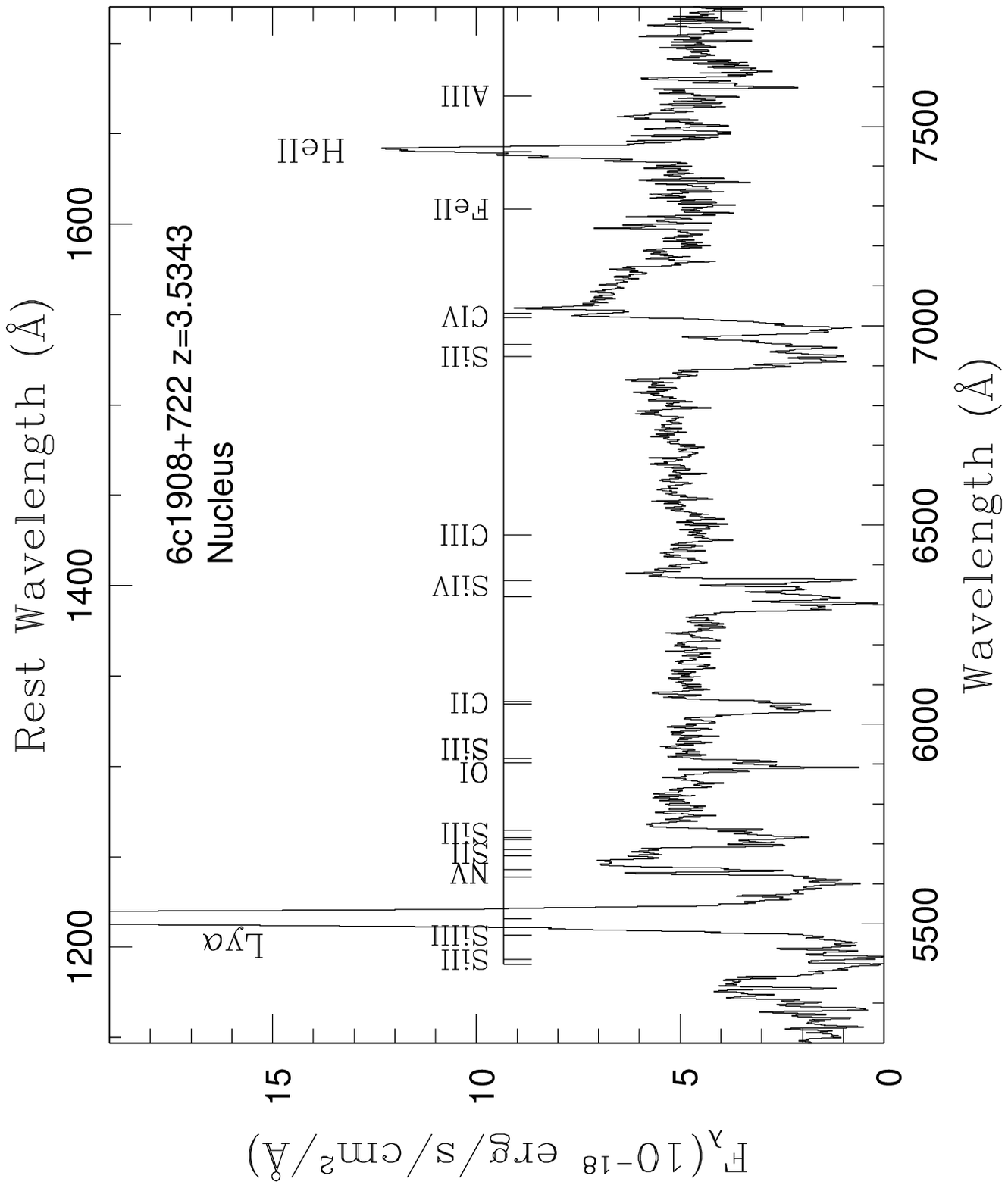}{2.4in}{-90}{42}{42}{-240}{241}
\vspace{-2in}
\plotfiddle{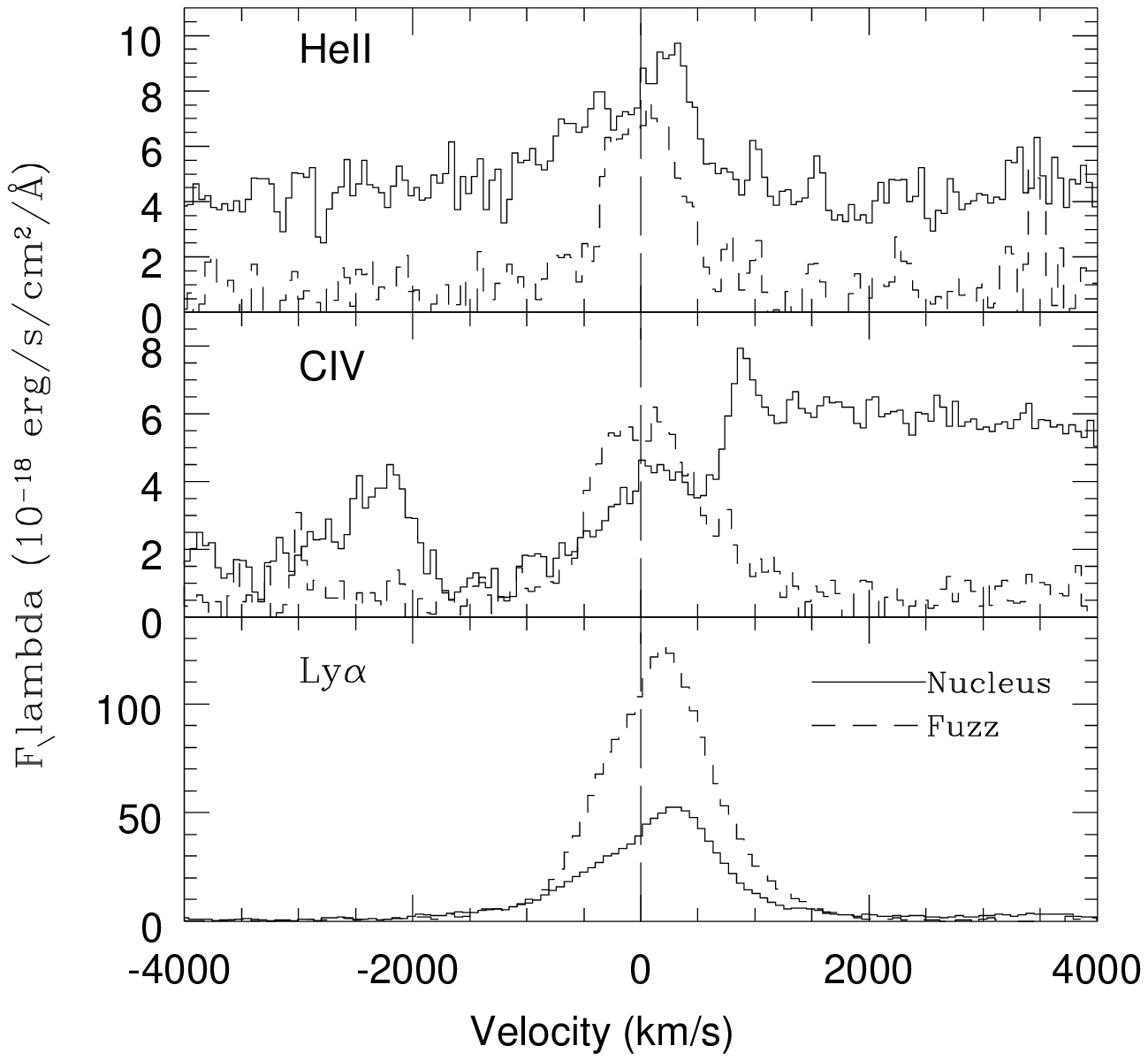}{2.4in}{0}{42}{42}{-25}{-25}
\vspace{-1in}
\caption{The left panel shows the spectrum of the $z=3.53$ radio galaxy 
6C~1908+722. The right panel shows the  
Ly$\alpha$, CIV and HeII line profiles. Note that
the profiles appear to be asymmetric both on and off the nucleus.}
\label{6c1908}
\end{figure}

Outflow phenomena have now been observed in nearly all $z>2$ radio
galaxies for which high-resolution, high signal-to-noise ratio data
exist. 3C256 shows an absorption feature in its spectrum, most likely
CIV$\lambda$1550 absorption arising in gas moving at a velocity
of 14300\kms\ relative to the radio galaxy (Dey \etal 1996).  Absorbing
gas associated with lower velocity outflows has now been observed in
several HzRGs, almost all of which show 
asymmetric \lya\ profiles suggesting the presence of
blue-shifted absorbing gas which appears to be spatially extended over
the entire emission line region (figure~\ref{lyapanels}; see also van Ojik 
\etal 1997).  One of the most spectacular cases is shown
in figure~\ref{6c1908}, where the absorption is likely to be spatially
extended, modifying the off-nuclear emission line profiles.
In the cases for which the data exist, the outflows extend over tens 
of kpc, and exhibit equivalent widths of $\sim 10$~\AA\ and 
FWHM$\simgt 2000$~\kms. In many cases, P Cygni -- like features are 
observed in both low- and high-ionization lines suggesting that the 
outflowing material has a wide range of ionization.

Evidence for outflows is also observed in the more normal (\ie,
non-AGN) $z\sim 3$ star-forming galaxies. For example, 0000-263~D6, a
`normal' (\ie, non-AGN) star-forming galaxy at $z=2.96$, also shows
P-Cygni like features associated with both the low-ionization and
high-ionization lines, perhaps suggesting the presence of a galaxian
wind (\eg, Spinrad \etal 1998).  If the outflow phenomena are indeed
ubiquitous, could they be a fundamental part of the AGN / galaxy
formation process? A telling analogy may be made to the low mass
star--formation process, where large--scale outflows are thought to
provide a mechanism for angular momentum loss, which is necessary to
enable accretion.

\section{Conclusions}

I have attempted to present an overview of our observations of $z>1$ radio
galaxies and describe our present observational understanding of the evolution
of the different components in these objects. The preliminary evidence
suggests that the AGN contribution to the UV light is less important at
$z\sim 4$ than at $z\sim 1$ and may reflect the spectral evolution of
the stelar component which dominates the UV light at higher redshifts.
Indeed, the present data, albeit sparse, qualitatively 
supports an evolutionary scenario in which powerful radio galaxies form 
the bulk of their stars before $z\sim 3.5-4$, and then evolve relatively 
quiescently (\ie, with little or no continuing star-formation) to $z\simlt1$.

This subject remains in its infancy and is still photon-starved. The
spectroscopic and polarimetric observations that are necessary to
elucidate the content and evolution of HzRGs
require long exposures on the largest telescopes, and our
understanding will therefore improve considerably during the next
decade with the availability of sensitive instruments on the Keck, 
VLT and Gemini telescopes.

\begin{acknow}
I thank my collaborators Wil van Breugel, Hy Spinrad, Bill Vacca,
Daniel Stern, Ski Antonucci, Mark Dickinson, Andrea Cimatti, Andrew
Bunker and Huub Rottgering for permitting me to present some of our
results prior to publication.  Much of the data presented here were
obtained at the W.~M.~Keck Observatories, and I thank the Observatory
staff for their expert assistance.  I am very grateful to the KNAW and
to NOAO for financial support that made it possible for me to attend
the conference. In particular, I thank Huub R\"ottgering, Philip Best,
Matt Lehnert and George Miley for providing me with the opportunity to
visit Amsterdam, for a stimulating conference, and particularly for
their extreme patience in waiting for this contribution.  
\end{acknow}

\end{document}